\documentclass[twocolumn,trackchanges]{aastex701}

\usepackage{graphicx}
\usepackage{float}    
\usepackage{placeins}
\usepackage{newtxtext,newtxmath,natbib}
\usepackage[T1]{fontenc}
\usepackage{appendix}
\usepackage{ae,aecompl}
\usepackage{hyperref}
\usepackage[svgnames]{xcolor}
\usepackage{tablefootnote}
 \usepackage{threeparttable}
\hypersetup{
colorlinks   = true, 
urlcolor     = blue, 
linkcolor    = blue, 
citecolor   = blue 
}

\definecolor{N}{RGB}{220,20,60}
\definecolor{N1}{RGB}{148,0,211}

\shorttitle{An Empirical Effective-Temperature Calibrations for Galactic B/A Supergiants} 
\shortauthors{N.~Vaidman et al.}

\begin{document}

\title{An Empirical Effective-Temperature Calibrations for Galactic B/A Supergiants}

\author[0000-0002-7449-0108]{N.~L.~Vaidman}
\affiliation{Fesenkov Astrophysical Institute, Observatory, 23, Almaty, 050020, Kazakhstan}
\affiliation{Al-Farabi Kazakh National University, Al-Farabi Ave., 71, 050040, Almaty, Kazakhstan}
\email{nva1dmann@gmail.com}

\author[0000-0002-3833-1038]{A.~S.~Miroshnichenko}
\affiliation{Fesenkov Astrophysical Institute, Observatory, 23, Almaty, 050020, Kazakhstan}
\affiliation{Al-Farabi Kazakh National University, Al-Farabi Ave., 71, 050040, Almaty, Kazakhstan}
\email{a_mirosh@uncg.edu}
\affiliation{Department of Physics and Astronomy, University of North Carolina at Greensboro, P.O. Box 26170, Greensboro, NC 27402-6170, USA}

\author[0009-0005-6851-7270]{Sh.~T.~Nurmakhametova}
\affiliation{Al-Farabi Kazakh National University, Al-Farabi Ave., 71, 050040, Almaty, Kazakhstan}
\email{shahidanurmahametova@gmail.com}

\author[0000-0001-8452-4220]{S.~Danford}
\affiliation{Department of Physics and Astronomy, University of North Carolina at Greensboro, P.O. Box 26170, Greensboro, NC 27402-6170, USA}
\email{danford@uncg.edu}

\author[0000-0001-5163-508X]{S.~A.~Khokhlov}
\affiliation{Fesenkov Astrophysical Institute, Observatory, 23, Almaty, 050020, Kazakhstan}
\affiliation{Al-Farabi Kazakh National University, Al-Farabi Ave., 71, 050040, Almaty, Kazakhstan}
\email{skhokh88@gmail.com}

\author[0000-0001-9878-0989]{I.~M.~Izmailova}
\affiliation{Fesenkov Astrophysical Institute, Observatory, 23, Almaty, 050020, Kazakhstan}
\email{chingis.omarov@gmail.com}

\author[0000-0002-1672-894X]{Ch.~T.~Omarov}
\affiliation{Fesenkov Astrophysical Institute, Observatory, 23, Almaty, 050020, Kazakhstan}
\email{chingis.omarov@gmail.com}

\author[0000-0003-2526-2683]{S.~V.~Zharikov}
\affiliation{Universidad Nacional Aut\'{o}noma de M\'{e}xico, Instituto de Astronom\'{i}a, AP 106,  Ensenada 22800, BC, M\'{e}xico}
\email{zhariaunam@gmail.com}

\author[0009-0008-3989-874X]{D.~E.~Yakhiya}
\affiliation{Al-Farabi Kazakh National University, Al-Farabi Ave., 71, 050040, Almaty, Kazakhstan}
\email{ddaniyaledu@gmail.com}

\author[0000-0001-9788-7485]{A.~T. Agishev}
\affiliation{Fesenkov Astrophysical Institute, Observatory, 23, Almaty, 050020, Kazakhstan}
\affil{Al-Farabi Kazakh National University, Al-Farabi Ave., 71, 050040, Almaty, Kazakhstan}
\email{aldiyar.agishev@gmail.com}

\author[0000-0001-6987-9058]{A.~A.~Khokhlov}
\affiliation{Al-Farabi Kazakh National University, Al-Farabi Ave., 71, 050040, Almaty, Kazakhstan}
\email{kh.azamat92@gmail.com}

\author[0009-0008-7289-1347]{D.~T.~Agishev}
\affiliation{Al-Farabi Kazakh National University, Al-Farabi Ave., 71, 050040, Almaty, Kazakhstan}
\email{agishev.pluto@gmail.com}

\begin{abstract}
We present empirical effective temperature ($T_{\rm eff}$) calibrations for Galactic supergiants of spectral types B5--A5 based on optical spectra. The relationships were derived from a reference sample with adopted literature temperatures and use equivalent widths, central line depths, and ratios of selected spectral features as temperature indicators. Quadratic relationships are established for individual diagnostics and their ratios. Collectively, the calibrations span a T$_{\rm eff}$ interval from 8\,400~K to 14\,700~K, while the validity range of each relationships is individually specified. Detailed quantitative atmospheric analyzes remain indispensable for deriving physically consistent stellar parameters. However, their application to extensive spectroscopic samples is observationally and computationally demanding. The empirical relationships presented here provide a homogeneous and readily applicable temperature scale for the characterization of Galactic BA supergiants. The complete calibration tables, including fitted coefficients and uncertainty information, together with an interactive Python tool to apply the relationships, are publicly available through Zenodo. The reduced continuum-normalized spectra and associated metadata are published through a VO--compliant service of the Kazakhstan National Virtual Observatory.
\end{abstract}


\section{Introduction}

Supergiants of B and A spectral types are among the visually brightest stars and can be observed spectroscopically at large distances. Their intrinsic luminosities and rich optical spectra make them important probes of stellar evolution and stellar atmosphere physics, while quantitative analysis of these objects also provides information on chemical composition, interstellar extinction, and distances in their host galaxies \citep{2017IAUS..329..297U,2012Ap&SS.341..131K}. 

In the Milky Way, BA-type supergiants are especially valuable because they can be studied in much greater spectroscopic detail and therefore serve as benchmarks for the interpretation of luminous blue stars in more distant systems.
However, they are not simple objects for quantitative spectroscopy. Their atmospheres are extended, departures from the local thermodynamic equilibrium are important, and in the more luminous objects stellar winds may affect important diagnostic features \citep{2006A&A...445.1099P,2006A&A...446..279C}. 

\begin{deluxetable*}{lrlccrr}
\tablecaption{Spectroscopic observations of the supergiant sample.\label{tab:obs}}
\tablewidth{0pt}
\tablehead{
\colhead{Observatory} &
\colhead{Dates} &
\colhead{Telescope / Spectrograph} &
\colhead{$R$} &
\colhead{Range} &
\colhead{Stars} &
\colhead{Spectra} \\
\colhead{} &
\colhead{} &
\colhead{} &
\colhead{} &
\colhead{[\AA]} &
\colhead{} &
\colhead{}
}
\startdata
TCO      & 2011--2026 & 0.81\,m / eShel    & 12\,000 & 3800--7800  & 135 & 1075 \\
ATO      & 2026       & 1.0\,m /  eShel    & 12\,000 & 3800--7800  &   3 &    3 \\
OAN SPM  & 2013--2019 & 2.1\,m /  REOSC    & 18\,000 & 3700--7800  &  46 &   50 \\
CFHT     & 2011--2015 & 3.6\,m /  ESPaDOnS & 65\,000 & 3760--8900  &   3 &    4 \\
\enddata

\tablecomments{
TCO -- Three College Observatory, North Carolina, USA;
ATO -- Assy-Turgen Observatory, Kazakhstan;
OAN SPM -- Observatorio Astron\'omico Nacional at San Pedro M\'artir, Baja California, Mexico;
CFHT -- Canada--France--Hawaii Telescope, Hawaii, USA.
}
\end{deluxetable*}

Studies of Galactic BA supergiants were typically based on relatively small groups (20--40 objects), whose $T_{\rm eff}$ alone or together with luminosities were determined by comparison with theoretical spectra computed with codes of various sophistication. For example, \citet{1995ApJS...99..659V} determined the effective temperatures $T_{\rm eff}$ and surface gravities $\log g$ by an LTE analysis of high-resolution spectra of 22 A0--F0 supergiants. This author also calculated absolute bolometric magnitudes from the objects' spectral type that turned out to be very different from early estimates based on kinematic distances \citep[e.g.,][]{1992AandAS...94..211G}.
Another example is a study by \citet{1999AandA...346..819V}, where fundamental parameters (e.g., $T_{\rm eff}$, luminosities, radii, and projected rotational velocities ($v\,\sin i$)) were derived for 31 bright A0--A5 supergiants using the {\it ATLAS9} code \citep{1979ApJS...40....1K}. 

More recent accurate determinations of the $T_{\rm eff}$ and $\log g$ rely on detailed non-LTE analyzes using together with Stark-broadened hydrogen lines and, where appropriate, helium lines \citep{2006A&A...445.1099P,2012AA...543A..80F}. These approaches provide physically consistent stellar parameters with high precision, but they are observationally and computationally demanding and not always optimal for the rapid analysis of large spectroscopic samples.

Several alternative parameter scales and semi-empirical approaches have been developed for hot luminous stars. In particular, homogeneous analyzes of Galactic BA supergiants have led to improved relationships between spectral type and $T_{\rm eff}$, as well as to photometric calibrations for these stars \citep{2012AA...543A..80F}. For B-type supergiants, $T_{\rm eff}$ calibrations based on the Balmer discontinuity in the BCD system have also been established \citep{2009AandA...501..297Z}. These methods are well motivated and powerful, especially when suitable spectrophotometric material is available, but they are conceptually different from compact empirical calibrations based directly on equivalent widths (EW) or EW ratios measured in optical spectra.

For cooler luminous stars, empirical $T_{\rm eff}$ diagnostics based on relative line strengths have proved highly successful. For Cepheids and F-G supergiants, \citet{2000A&A...358..587K} derived analytical relationships between $T_{\rm eff}$ and selected line-depth ratios, showing that carefully chosen pairs of spectral lines can provide very precise temperature estimates. This approach was later extended to a large sample of FGK supergiants by \citet{2007MNRAS.378..617K}. For hotter supergiants, there is still no comparable homogeneous set of widely used empirical $T_{\rm eff}$ calibrations based on various properties of optical spectral lines, such as EWs, intensity, or intensity ratios.

This calibration gap is important for several reasons. First, large spectroscopic collections of Galactic BA supergiants are now available, and a rapid but internally consistent estimate of $T_{\rm eff}$ is often desirable at the initial stages of astrophysical analysis. Second, empirical relationships of this kind can provide useful starting values for more sophisticated non-LTE modeling. Third, they are particularly attractive when dealing with spectra of moderate resolution, incomplete wavelength coverage, or variable data quality, where a first-order $T_{\rm eff}$ estimate is needed before a full atmospheric analysis is attempted. The existence of calibration stars with accurately determined atmospheric parameters from modern quantitative studies makes this approach feasible and timely \citep{2006A&A...445.1099P,2012AA...543A..80F}.

\begin{figure*}[!t]
\centering
\includegraphics[width=1\linewidth]{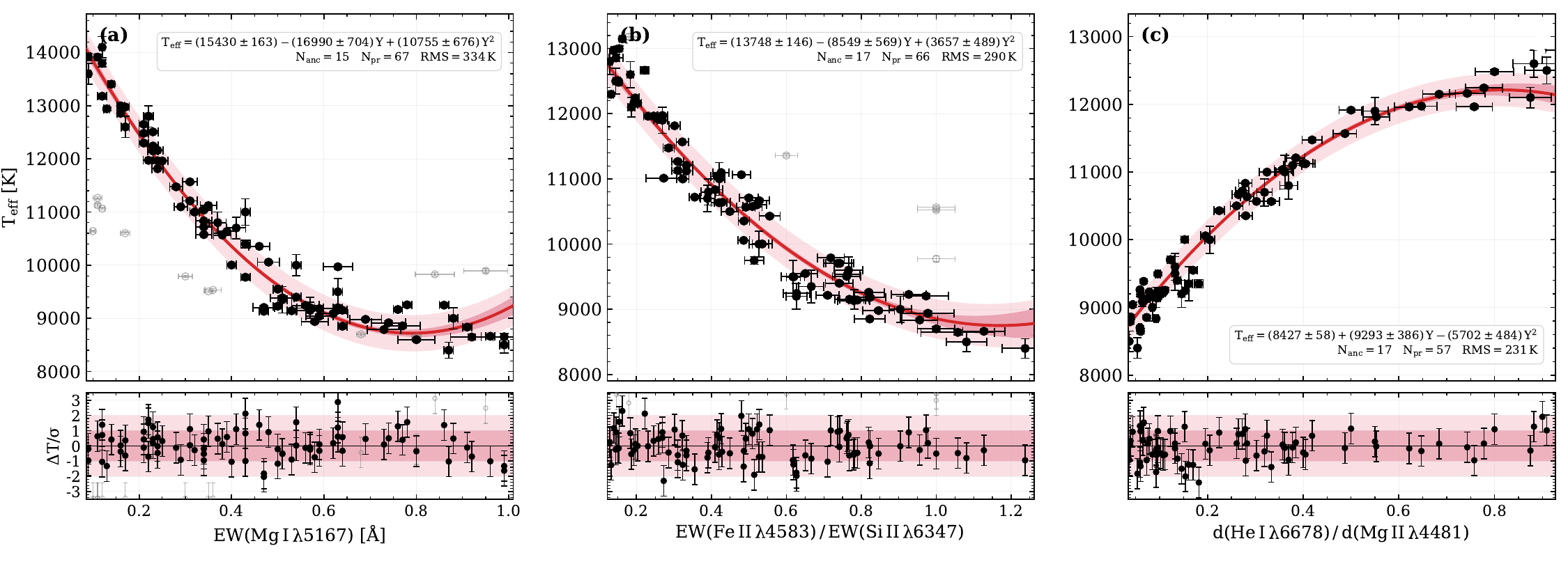} 
\caption{
Examples of empirical $T_{\rm eff}$ calibrations based on
(a) the equivalent width of Mg\,\textsc{i}~$\lambda5167$,
(b) the equivalent-width ratio
Fe\,\textsc{ii}~$\lambda4583$/Si\,\textsc{ii}~$\lambda6347$,
and (c) the line-depth ratio
He\,\textsc{i}~$\lambda6678$/Mg\,\textsc{ii}~$\lambda4481$.
Filled black symbols represent the reference-star measurements used to
derive the calibration relationships. Open gray symbols, where shown,
represent program star measurements and are displayed only to
illustrate the subsequent application of the relationships; they were not
used in the fitting procedure.
The solid curves show the best-fitting quadratic relationships derived from
the reference-star sample. In the upper panels, the darker and lighter
shaded regions indicate the propagated uncertainty of the fitted
relationships and the total calibration uncertainty including the RMS scatter,
respectively. The lower panels show the normalized residuals,
$\Delta T/\sigma$; the darker and lighter shaded bands indicate the
$\pm1\sigma$ and $\pm2\sigma$ intervals, respectively. The fitted
equations, the numbers of reference stars used in the fits, and the RMS
scatter are given in the corresponding panels.
}
\label{fig:calib_EW}
\end{figure*}

In this paper, we use a sample of bright Galactic B5--A5 supergiants with well-established $T_{\rm eff}$ as calibration standards to derive empirical relationships between $T_{\rm eff}$ and select EW and EW ratios of spectral lines in the optical region. The adopted diagnostics are based on temperature-sensitive lines that remain measurable across parts of the chosen range of $T_{\rm eff}$ in spectra of sufficient quality. 

Our aim is not to replace detailed quantitative spectroscopy, but to provide a practical empirical framework that can be applied rapidly and consistently to a much larger observational sample. In this way, the present study seeks to bridge the gap between physically comprehensive model atmosphere analysis tools and efficient empirical tools for the $T_{\rm eff}$ characterization of {\bf hot} Galactic supergiants.

\section{Sample Selection and Observations}\label{sec:obs}

The working sample includes bright stars that were classified as B5 -- A5 supergiants (luminosity types {\sc ia}, {\sc iab}, or {\sc ib}) or bright giants (luminosity type {\sc ii}) down to a visual magnitude $V \sim 9$ mag observable from latitudes $\ge 30^{\circ}$ in the Northern hemisphere, where most observatories listed in Table\,\ref{tab:obs}, are located.

The MK types for the stars were taken from a wide range of sources (e.g., SIMBAD and various spectroscopic surveys) that were available before the start of our project in 2011. Stars with different assessments of their MK types were kept in the sample, if one of them indicated a high luminosity. This requirement was set to independently derive the fundamental parameters of such stars.

The project was intended to be done mainly at a medium spectral resolution ($R \sim 10,000 - 20,000$) and $1-2$\,meter class telescopes to allow for a large range of brightness with a little need for observing time requests.

The observational material was assembled from optical spectra of Galactic B and A-type supergiants obtained at different facilities and during different observing campaigns. The data set includes spectra with different spectral resolution, wavelength coverage, and number of available epochs per star. This heterogeneous material was used to increase the sample size and provide a broad coverage of the optical diagnostic lines required for the measurement of line-strength indicators.

Most stars were observed at the 0.81\,m telescope of the Three College Observatory (TCO) in central North Carolina, while a fraction of the faintest stars ($V \ge 7$ mag) were observed at the 2.1\,m telescope of the Observatorio Astron\'omicao Nacional San Pedro Martir (OAN SPM) in Baja California, Mexico. Several spectra were taken at the 1\,m telescope of the Assy-Turgen Observatory (ATO) located near Almaty in southern Kazakhstan and the 3.6\,m Canada-France-Hawaii Telescope (CFHT) located at the summit of Mauna Kea, Big Island, Hawaii, USA to complement the collection.

A summary of the spectroscopic material is given in Table~\ref{tab:obs}. Each sample star was observed at least 2--3 times (the faintest stars with $V > 8.0$ mag). However, brighter stars were typically observed 4--7 times on different years. Also, several stars previously suspected of showing spectral line variability were observed 30--200 times (e.g., $\alpha$ Cyg, $\beta$ Ori, $\sigma$ Cyg, HD\,21389, HD\,223960) to study the variations separately from this study. Since the spectra originate from different instrumental setups, all line parameters were measured with the same homogeneous procedure described in Sect.~\ref{sec:line_measurements}.

The TCO spectra were reduced using the IRAF \texttt{echelle} package, including bias subtraction, extraction of spectral orders, wavelength calibration with ThAr lamp exposures, and co-addition of individual frames. No flat-field correction was applied due to the low pixel-to-pixel sensitivity variations of the detector \citep{2023Galax..11....8M}. The same facility and general \'echelle-spectroscopic approach have previously been used for studies of emission-line and interacting stellar systems \citep{2020CoSka..50..513M,2022ApJ...936..129N, 2025Galax..13..101V}. 

The ATO and SPM spectra were reduced with standard IRAF echelle routines, including bias subtraction, flat-field correction, extraction of echelle orders, wavelength calibration using ThAr comparison spectra, continuum normalization, and heliocentric correction.

The ESPaDOnS spectra were reduced with the Libre-ESpRIT pipeline \citep{1997MNRAS.291..658D}, which performs bias and dark subtraction, flat-field correction, optimal extraction of \'echelle orders, wavelength calibration using ThAr spectra, sky subtraction, and heliocentric correction.

\begin{deluxetable*}{l l l l @{\hspace{0.8cm}} l l l l}
\tabletypesize{\scriptsize}
\tablecaption{Calibration stars used to build the empirical $T_{\rm eff}$ calibration. \label{tab:calib_stars}}
\tablewidth{0pt}
\tablehead{
\colhead{No.} & \colhead{Star} & \colhead{$T_{\rm eff,lit}$} & \colhead{Ref} &
\colhead{No.} & \colhead{Star} & \colhead{$T_{\rm eff,lit}$} & \colhead{Ref} \\
\colhead{} & \colhead{} & \colhead{(K)} & \colhead{} &
\colhead{} & \colhead{} & \colhead{(K)} & \colhead{}
}
\startdata
1 & BD+60 2582 & $11 900\pm200$ & 1 & 18 & HD 164865 & $10 500$ & 4 \\
2 & HD 7902 & $15 500$ & 6 & 19 & HD 165784 & $9000\pm200$ & 2 \\
3 & HD 12301 & $12 600\pm200$ & 1 & 20 & HD 183143 & $12 800\pm200$ & 3 \\
4 & HD 12953 & $9200\pm200$ & 2 & 21 & HD 184943 & $11 900$ & 3 \\
5 & HD 13476 & $8500\pm150$ & 2 & 22 & HD 186745 & $12 500\pm200$ & 2 \\
6 & HD 13744 & $9500\pm250$ & 1 & 23 & HD 187982 & $9300\pm250$ & 2 \\
7 & HD 14433 & $9150\pm150$ & 2 & 24 & HD 191243 & $14 000\pm300$ & 3 \\
8 & HD 14489 & $9350\pm250$ & 2 & 25 & HD 195324 & $9200\pm150$ & 2 \\
9 & HD 20041 & $10 000\pm200$ & 2 & 26 & HD 199478 & $12 700\pm200$ & 3 \\
10 & HD 21291 & $10 800\pm200$ & 2 & 27 & HD 202850 & $10 800\pm200$ & 5 \\
11 & HD 25914 & $13 600\pm200$ & 3 & 28 & HD 207673 & $9250\pm100$ & 2 \\
12 & HD 34085 & $12 100\pm150$ & 2 & 29 & HD 208501 & $12 700\pm200$ & 2 \\
13 & HD 36371 & $14 600\pm300$ & 3 & 30 & HD 210221 & $8400\pm150$ & 2 \\
14 & HD 39970 & $10 300\pm200$ & 2 & 31 & HD 212593 & $11 200$ & 1 \\
15 & HD 46300 & $10 000\pm200$ & 1 & 32 & HD 213470 & $8400\pm150$ & 2 \\
16 & HD 87737 & $9600\pm200$ & 7 & 33 & HD 223960 & $10 700\pm200$ & 2 \\
17 & HD 164353 & $15 500\pm1000$ & 2 &  &  &  &  \\
\enddata
\tablecomments{References: (1)~\citep{2012AA...543A..80F}; (2)~\citep{2021AA...650A.128G}; (3)~\citep{2022AandA...668A..92W}; (4)~\citep{1988AAS...72..259D}; (5)~\citep{2010AandA...517A..38P}; (6)~\citep{1999AA...349..553M}; (7)~\citep{1995ApJS...99..659V}.}
\end{deluxetable*}

The reference sample consists exclusively of Galactic BA-type supergiants. Although homogeneous abundance determinations are not available for all calibration stars, detailed quantitative analyses of Galactic BA- and B-type supergiants generally indicate near-solar heavy-element abundances \citep{2006A&A...445.1099P,2022AandA...668A..92W}. The empirical relationships derived here should therefore be regarded as representative primarily of the approximately solar-metallicity regime of Galactic BA supergiants. Metallicity was not included as an independent parameter in the construction of the calibrations.

The compiled spectroscopic database, which spans over a decade of observations, represents a homogeneous core of medium-resolution data.
Initial results of the current project were reported in \citet{2013msao.confE.169M}.
To ensure long-term preservation and enable efficient data mining, fully reduced continuum-normalized spectra have been integrated into the infrastructure of the Kazakhstan National Virtual Observatory (KazVO)\footnote{\url{https://vo.fai.kz/}}. The sample metadata are structured in strict compliance with the IVOA Observation Core Components (ObsCore) data model and support queries via the Simple Spectral Access Protocol (SSAP). A more detailed description of the technical architecture of the public archive is provided in Sect.~\ref{sec:vo_service}.

\section{Selection of diagnostic lines}
\label{sec:line_selection}

The initial diagnostic line list was constructed from optical transitions that are commonly used in the classification and quantitative spectroscopy of early-type luminous stars. Hydrogen Balmer lines were included because Stark-broadened hydrogen profiles are standard constraints in the determination of atmospheric parameters of BA- and B-type supergiants. Helium lines and metal ionization equilibria were included because they provide temperature-sensitive diagnostics throughout the hot-supergiant regime \citep{2006A&A...445.1099P,2012AA...543A..80F,2022AandA...668A..92W}.

The He \,\textsc{i}~$\lambda4471$ and Mg \,\textsc{ii}~$\lambda4481$ lines were included as a classical late-B / early-A diagnostic pair: He \,\textsc{i} weakens towards the A-type domain, whereas Mg\,\textsc{ii}~$\lambda4481$ remains prominent and provides a useful contrast in this transition region \citep{2009ssc..book.....G}. Additional He\,\textsc{i} lines were retained for the hotter part of the sample, while He\,\textsc{ii} lines were included only as diagnostics for the hottest objects and were not required to be measurable throughout the full BA-supergiant range.

Silicon lines were selected in several ionization stages, Si\,\textsc{ii}, Si\,\textsc{iii}, and Si\,\textsc{iv}, because ionization equilibria are among the primary temperature indicators in quantitative analyses of hot luminous stars \citep{2006A&A...445.1099P,2022AandA...668A..92W}. Mg\,\textsc{i} and low-ionization metallic lines, including Fe\,\textsc{ii}, Ti\,\textsc{ii}, Cr\,\textsc{ii}, and Fe\,\textsc{i}, were included because they provide numerous measurable features in late-B and A supergiants and allow empirical line-strength indicators to be constructed over a broad optical wavelength range.

Additional C\,\textsc{ii}, N\,\textsc{ii}, O\,\textsc{ii}, Sc\,\textsc{ii}, Sr\,\textsc{ii}, and Ca\,\textsc{ii} features were considered as auxiliary diagnostics where they were covered by the spectra and not affected by severe blending, interstellar absorption, or poor continuum definition. The use of line strengths and line-strength ratios as empirical temperature indicators follows the same general principle that lines with different temperature sensitivity change differently with \(T_{\rm eff}\) \citep{2007MNRAS.378..617K}.

This list was used as an initial measurement pool rather than as a fixed set of final diagnostics. A line was retained for quantitative use only if it was covered by the available spectra, measurable at the instrumental resolution, and passed the quality-control procedure described in Sect.~\ref{sec:line_measurements}.

\section{Measurement of spectral-line parameters}
\label{sec:line_measurements}

Spectral-line parameters were measured with a custom Python-based code developed for homogeneous analysis of spectra obtained with different instrumental setups. The code operates on one-dimensional continuum-normalized spectra and performs local re-normalization around each selected feature.

For each line, a wavelength window was defined around the expected line position, using the laboratory wavelength as the reference and allowing for the stellar radial velocity when applicable. The local continuum was determined from adjacent line-free regions. Regions affected by strong blends, broad Balmer wings, telluric residuals, or normalization defects were excluded from the continuum estimate.

Line profiles were fitted with analytic absorption profiles to determine the line center, depth, FWHM, and measurement boundaries. The code allows both single-component and multi-component profile representations, making it applicable to simple narrow lines as well as complex, blended, or broad features. The fitted profile was used to define a stable line window, while the EW was measured from the observed locally normalized spectrum rather than from the fitted model.

Measurements were accepted only when the continuum placement was reliable, the fitted center was consistent with the expected line position, the fit converged to a stable solution, and no severe blending or telluric contamination was present. For accepted features, the final output includes EW, line depth, line center, and FWHM. EW ratios were computed only from lines that passed the same quality-control procedure.

Since the calibrations are based on line parameters measured from continuum-normalized spectra, they are largely insensitive to 
interstellar extinction, which mostly affects the observed continuum.
Their application to reddened stars therefore remains appropriate, provided that sufficient signal-to-noise ratio is achieved and that the diagnostic features are not
blended with diffuse interstellar bands.

\section{Construction of empirical temperature calibrations}
\label{sec:teff_calibrations}

The construction of the empirical calibrations was performed in two stages. First, an initial set of relationships was derived using only the reference stars with $T_{\rm eff}$ adopted from the literature. These stars define the external temperature scale of the calibration. For each reference star, the measured spectral-line parameters were converted into empirical indicators \(Y\), including individual EW, individual line depths, EW ratios, line-depth ratios, and ratios involving different ionization stages of the same element.

For each candidate indicator, the adopted reference temperatures were fitted with a quadratic relationship,
\begin{equation}
T_{\rm eff}=A+B\,Y+C\,Y^2 .
\label{eq:teff_calibration}
\end{equation}

The quadratic form was adopted as a common low-order approximation for all retained indicators. It provides the minimum flexibility required to reproduce the main curvature of the empirical relationships over the observed calibration interval while preserving a uniform functional form for EW-, line-depth-, and ratio-based diagnostics. Higher-order polynomials were not considered because the behavior near the boundaries of the calibration sample is less tightly constrained and may become excessively sensitive to individual measurements. The adopted relationships are therefore used only within the corresponding calibrated ranges of $Y$ and $T_{\rm eff}$.

The coefficients were obtained by least-squares fitting, and their formal uncertainties were derived from the covariance matrix of the fit \citep{2003drea.book.....B}.

The empirical relationships were derived exclusively from the reference stars with adopted literature $T_{\rm eff}$. Thus, the fitted coefficients, covariance matrices, RMS residuals, and calibrated validity ranges were determined solely by the reference-star sample.

The resulting relationships were then applied to the program stars. For each object, every available diagnostic provided an independent estimate of \(T_{\rm eff}\), provided that the measured value of \(Y\) was within the calibrated range of the corresponding relationship. The uncertainty of each individual temperature estimate was computed as described below. 

A robust consistency filtering step was applied separately to the set of individual temperature estimates obtained for each program star. Outlying estimates were rejected using a median absolute deviation criterion with a threshold of \(3\,\sigma_{\rm MAD}\), where the MAD was scaled by 1.4826 to be comparable to the standard deviation for a normal distribution \citep{1974JASA...69..383H}. Diagnostics that produced recurrently discrepant estimates for at least 30\% of the program stars were flagged as operationally unstable and were not used in the final average. This application-stage filtering did not modify the fitted coefficients, covariance matrices, RMS residuals, or validity ranges of the reference-star calibrations.

Program star measurements were therefore used only to apply the calibrations and assess their practical performance; they were not included in the derivation or refitting of the empirical relationships.

\begin{deluxetable*}{ccccrrrrr}
\tabletypesize{\scriptsize}
\tablecaption{Selected EW-based empirical temperature calibrations.
\label{tab:ew_calibrations}}
\tablehead{
\colhead{$\lambda_1$} &
\colhead{El$_1$} &
\colhead{$\lambda_2$} &
\colhead{El$_2$} &
\colhead{RMS} &
\colhead{$T_{\rm eff}$ range} &
\colhead{$A$} &
\colhead{$B$} &
\colhead{$C$} \\
\colhead{(\AA)} &
\colhead{} &
\colhead{(\AA)} &
\colhead{} &
\colhead{(K)} &
\colhead{(K)} &
\colhead{} &
\colhead{} &
\colhead{}
}
\startdata
4583 & Fe\,\textsc{ii} & --   & --              & 290 & 8400-12800 & $14064\pm147$ & $-16295\pm899$ & $12376\pm1246$ \\
4634 & Cr\,\textsc{ii} & --   & --              & 350 & 8400-12600 & $12562\pm118$ & $-24408\pm1611$ & $39577\pm4672$ \\
4178 & Fe\,\textsc{ii} & --   & --              & 300 & 8400-12800 & $12747\pm94$  & $-15174\pm729$ & $14762\pm1168$ \\
4824 & Cr\,\textsc{ii} & --   & --              & 280 & 8400-11000 & $11682\pm175$ & $-22928\pm2541$ & $47286\pm8132$ \\
5167 & Mg\,\textsc{i}  & --   & --              & 330 & 8400-14100 & $15430\pm163$ & $-16990\pm704$ & $10755\pm676$  \\
5875 & He\,\textsc{i}  & --   & --              & 400 & 8500-14700 & $8439\pm78$   & $9352\pm504$   & $-4523\pm616$  \\
4178 & Fe\,\textsc{ii} & 4383 & Fe\,\textsc{i}  & 300 & 8800-12800 & $13364\pm666$ & $-5650\pm1951$ & $1371\pm1358$  \\
6678 & He\,\textsc{i}  & 4481 & Mg\,\textsc{ii} & 330 & 8400-12800 & $8449\pm94$   & $5306\pm337$   & $-1885\pm218$  \\
4583 & Fe\,\textsc{ii} & 4923 & Fe\,\textsc{ii} & 240 & 8400-12800 & $13103\pm193$ & $-3759\pm717$  & $-1021\pm618$  \\
5875 & He\,\textsc{i}  & 4481 & Mg\,\textsc{ii} & 240 & 8500-12800 & $8396\pm66$   & $4470\pm184$   & $-1331\pm99$   \\
4583 & Fe\,\textsc{ii} & 4921 & He\,\textsc{i}  & 260 & 8400-12800 & $13371\pm174$ & $-5087\pm593$  & $629\pm467$    \\
4583 & Fe\,\textsc{ii} & 6347 & Si\,\textsc{ii} & 290 & 8400-13100 & $13748\pm146$ & $-8549\pm569$  & $3657\pm489$   \\
4383 & Fe\,\textsc{i}  & 4416 & O\,\textsc{ii}  & 360 & 9000-12600 & $8366\pm183$  & $706\pm77$     & $-30\pm5$      \\
4583 & Fe\,\textsc{ii} & 4481 & Mg\,\textsc{ii} & 280 & 8400-12800 & $15431\pm248$ & $-10562\pm683$ & $4167\pm440$   \\
4634 & Cr\,\textsc{ii} & 4383 & Fe\,\textsc{i}  & 330 & 8800-12600 & $12879\pm236$ & $-8710\pm1293$ & $4537\pm1639$  \\
4383 & Fe\,\textsc{i}  & 4583 & Fe\,\textsc{ii} & 380 & 8800-12800 & $5607\pm685$  & $5239\pm1058$  & $-980\pm388$   \\
\enddata
\tablecomments{
The calibrations have the form \(T_{\rm eff}=A+B\,Y+C\,Y^2\).
For rows with no second line listed, \(Y=\mathrm{EW}(\lambda_1)\).
For line-ratio calibrations, \(Y=\mathrm{EW}(\lambda_1)/\mathrm{EW}(\lambda_2)\).
The quoted uncertainties of \(A\), \(B\), and \(C\) are formal uncertainties from the covariance matrix of the least-squares fit.
RMS is the root-mean-square scatter of the calibration residuals in \(T_{\rm eff}\).
}
\end{deluxetable*}

The complete set of calibration coefficients, coefficient uncertainties, covariance matrices, calibrated $Y$ ranges, and numbers of stars used in each relationship is provided in the machine-readable table. A stand-alone Python calculator that implements the final empirical calibrations based on EW and line-depth, including the full propagation of uncertainties according to Equation~(\ref{eq:teff_uncertainty}), is publicly available through Zenodo \citep{Vaidman2026Zenodo}.

\begin{deluxetable*}{ccccccccc}
\tabletypesize{\scriptsize}
\tablecaption{Selected line-depth empirical temperature calibrations.
\label{tab:depth_calibrations}}
\tablehead{
\colhead{$\lambda_1$} &
\colhead{El$_1$} &
\colhead{$\lambda_2$} &
\colhead{El$_2$} &
\colhead{RMS} &
\colhead{$T_{\rm eff}$ range} &
\colhead{$A$} &
\colhead{$B$} &
\colhead{$C$} \\
\colhead{(\AA)} &
\colhead{} &
\colhead{(\AA)} &
\colhead{} &
\colhead{(K)} &
\colhead{(K)} &
\colhead{} &
\colhead{} &
\colhead{}
}
\startdata
6678 & He\,\textsc{i}  & --   & --              & 355 & 8400--14700 & $8416\pm75$   & $19200\pm1104$  & $-16776\pm3072$ \\
4588 & Cr\,\textsc{ii} & --   & --              & 243 & 8400--12600 & $12574\pm79$  & $-27187\pm1256$ & $49705\pm4182$ \\
4824 & Cr\,\textsc{ii} & --   & --              & 239 & 8400--11000 & $11740\pm134$ & $-22411\pm1897$ & $42293\pm6153$ \\
4583 & Fe\,\textsc{ii} & --   & --              & 258 & 8400--13000 & $14204\pm156$ & $-20750\pm1298$ & $18833\pm2465$ \\
4178 & Fe\,\textsc{ii} & --   & --              & 292 & 8400--12800 & $13259\pm131$ & $-21731\pm1389$ & $26347\pm3264$ \\
5875 & He\,\textsc{i}  & --   & --              & 376 & 8500--14700 & $7953\pm135$  & $14097\pm1350$  & $-3344\pm2925$ \\
5167 & Mg\,\textsc{i}  & --   & --              & 342 & 8400--14100 & $15651\pm254$ & $-22724\pm1688$ & $18039\pm2582$ \\
4383 & Fe\,\textsc{i}  & 5055 & Si\,\textsc{ii} & 135 & 9000--12800 & $13975\pm220$ & $-7329\pm723$   & $2400\pm559$ \\
4383 & Fe\,\textsc{i}  & 5167 & Mg\,\textsc{i}  & 178 & 9000--12800 & $13971\pm555$ & $-5412\pm3049$  & $-5718\pm3992$ \\
4383 & Fe\,\textsc{i}  & 4481 & Mg\,\textsc{ii} & 200 & 8650--12800 & $13853\pm264$ & $-13064\pm1822$ & $5780\pm2959$ \\
4383 & Fe\,\textsc{i}  & 3970 & H$\epsilon$     & 220 & 9000--13100 & $13890\pm181$ & $-22549\pm1977$ & $25401\pm4786$ \\
4383 & Fe\,\textsc{i}  & 5041 & Si\,\textsc{ii} & 185 & 9000--12800 & $13952\pm238$ & $-5110\pm595$   & $1162\pm346$ \\
4824 & Cr\,\textsc{ii} & 4713 & He\,\textsc{i}  & 182 & 8800--11000 & $11229\pm113$ & $-741\pm87$     & $64\pm13$ \\
4383 & Fe\,\textsc{i}  & 6347 & Si\,\textsc{ii} & 199 & 9000--12800 & $13866\pm179$ & $-13668\pm1145$ & $9220\pm1656$ \\
4383 & Fe\,\textsc{i}  & 4026 & He\,\textsc{i}  & 258 & 8800--12800 & $12715\pm106$ & $-3059\pm205$   & $664\pm80$ \\
5875 & He\,\textsc{i}  & 4481 & Mg\,\textsc{ii} & 172 & 8500--12800 & $8038\pm70$   & $7080\pm303$    & $-2882\pm278$ \\
4383 & Fe\,\textsc{i}  & 4102 & H$\delta$       & 211 & 8800--13100 & $13861\pm146$ & $-20159\pm1521$ & $17980\pm3540$ \\
6678 & He\,\textsc{i}  & 4481 & Mg\,\textsc{ii} & 231 & 8400--12800 & $8427\pm58$   & $9293\pm386$    & $-5702\pm484$ \\
4588 & Cr\,\textsc{ii} & 4340 & H$\gamma$       & 228 & 8400--12600 & $13171\pm99$  & $-24052\pm1114$ & $33345\pm2724$ \\
4383 & Fe\,\textsc{i}  & 5169 & Fe\,\textsc{ii} & 181 & 8800--12800 & $13949\pm482$  & $-5394\pm2517$  & $-5645\pm3127$ \\
4383 & Fe\,\textsc{i}  & 5018 & Fe\,\textsc{ii} & 208 & 9000--12800 & $13947\pm721$  & $-4355\pm3548$  & $-5628\pm4201$ \\
4571 & Mg\,\textsc{i}  & 4481 & Mg\,\textsc{ii} & 285 & 8400--10400 & $11326\pm287$  & $-10903\pm2149$ & $12503\pm3753$ \\
4383 & Fe\,\textsc{i}  & 4923 & Fe\,\textsc{ii} & 249 & 9000--13100 & $13075\pm347$  & $-168\pm1907$   & $-11582\pm2467$ \\
4383 & Fe\,\textsc{i}  & 5316 & Fe\,\textsc{ii} & 374 & 8800--12800 & $15013\pm2275$ & $-4296\pm7357$  & $-4235\pm5854$ \\
4383 & Fe\,\textsc{i}  & 4233 & Fe\,\textsc{ii} & 374 & 8800--12800 & $12932\pm1543$ & $789\pm6648$    & $-11486\pm6992$ \\
5167 & Mg\,\textsc{i}  & 4481 & Mg\,\textsc{ii} & 395 & 8400--12800 & $22585\pm1317$ & $-26595\pm3716$ & $12662\pm2564$ \\
\enddata

\tablecomments{
The calibrations have the form \(T_{\rm eff}=A+B\,Y+C\,Y^2\).
For rows with no second line listed, \(Y=d(\lambda_1)\), where \(d\) is the measured line depth.
For line-ratio calibrations, \(Y=d(\lambda_1)/d(\lambda_2)\).
The quoted uncertainties of \(A\), \(B\), and \(C\) are formal uncertainties from the covariance matrix of the least-squares fit.
RMS is the root-mean-square scatter of the calibration residuals in \(T_{\rm eff}\).
The units of \(A\), \(B\), and \(C\) follow from the adopted definition of \(Y\).
The relationships should not be extrapolated outside the listed \(T_{\rm eff}\) ranges.
}
\end{deluxetable*}

\subsection{Uncertainty estimates}
\label{sec:uncertainties}

For each accepted calibration, the formal uncertainties of the coefficients \(A\), \(B\), and \(C\) in Equation~(\ref{eq:teff_calibration}) were obtained from the covariance matrix of the least-squares fit. These uncertainties
describe how well the empirical relationship is constrained by the reference-star sample and by the scatter of the calibration points.

When a calibration was applied to a star, the uncertainty of the resulting temperature estimate was computed by combining three terms: the propagated uncertainty of the measured indicator \(Y\), the uncertainty of the fitted
calibration curve, and the RMS scatter of the calibration residuals. This was calculated as
\begin{equation}
\sigma^2(T_{\rm eff}) =
\left[\left(B+2CY\right)\sigma_Y\right]^2
+
\mathbf{J}\,\mathbf{C}_{\rm coef}\,\mathbf{J}^{T}
+
{\rm RMS}^2 ,
\label{eq:teff_uncertainty}
\end{equation}
where \(\sigma_Y\) is the uncertainty of the measured indicator, \(\mathbf{J}=(1,Y,Y^2)\), and \(\mathbf{C}_{\rm coef}\) is the covariance matrix of the fitted coefficients.
For line-ratio indicators, \(\sigma_Y\) was propagated from the uncertainties of the two measured component lines. The final temperature of a program star was calculated as the inverse-variance weighted mean of all retained individual estimates.

To avoid unrealistically small formal errors when many indicators were available,
the final uncertainty was constrained by the scatter of the retained individual
estimates. We adopted
\begin{equation}
\sigma_{\mathrm{final}} =
\max(\sigma_{\mathrm{wmean}}, \sigma_{\mathrm{ind}}),
\label{eq:final_teff_uncertainty}
\end{equation}
where \(\sigma_{\mathrm{wmean}}\) is the uncertainty of the inverse-variance
weighted mean and \(\sigma_{\mathrm{ind}}\) is the standard deviation of the
individual temperature estimates retained after consistency filtering.

\section{Results}
\label{sec:results}

\subsection{Empirical calibration relationships}
\label{sec:calib_performance}

The final set of retained empirical calibrations is presented in Tables~\ref{tab:ew_calibrations} and~\ref{tab:depth_calibrations}. Table~\ref{tab:ew_calibrations} lists the EW based relationships, including both single-line indicators and EW ratios. For each diagnostic, the table gives the adopted line or paie of lines, the RMS scatter of the calibration residuals, the calibrated
\(T_{\rm eff}\) range, and the coefficients \(A\), \(B\), and \(C\) of Equation~(\ref{eq:teff_calibration}). Table~\ref{tab:depth_calibrations} has the same structure, but is based on line depths and line-depth ratios instead of EW.

The selected EW-based calibrations cover the temperature interval \(8400\)-\(14700\)~K, although the valid range differs from one diagnostic to another. Their RMS scatter is typically \(240\)-\(400\)~K. The smallest scatter among the selected EW relationships is obtained for several line-ratio diagnostics, including Fe\,\textsc{ii}~\(\lambda4583\)/Fe\,\textsc{ii}~\(\lambda4923\) and He\,\textsc{i}~\(\lambda5875\)/Mg\,\textsc{ii}~\(\lambda4481\), both with RMS values of about \(240\)~K. This confirms that ratios of lines with different temperature sensitivity can provide more stable empirical indicators than some individual EW.

The line-depth calibrations show a similar behavior, with RMS values ranging from about \(135\) to \(395\)~K. The best-performing depth-based relationships are mainly line-depth ratios, consistent with the general line-depth ratio approach in which spectral lines with different temperature sensitivities are combined to construct sensitive temperature indicators \citep{2000A&A...358..587K,2007MNRAS.378..617K}.

Single-line calibrations were also retained, but were interpreted more conservatively. They are useful for spectra with moderate resolution or limited wavelength coverage, where only a small number of diagnostic features can be measured reliably. In such cases, individual line strengths can still provide practical first-order temperature estimates, provided that the lines are sufficiently isolated, the continuum placement is stable, and the measurements remain within the calibrated range.

An important practical property of the present calibrations is that they do not require an explicit correction for interstellar reddening. The adopted temperature indicators are EWs and central line depths defined relative to the local stellar continuum.
The slowly changing with wavelength interstellar extinction affects the line and adjacent continuum essentially by the same multiplicative factor and therefore does not introduce a systematic change in the measured line strengths and EWs.
This makes the method intrinsically less sensitive to reddening than temperature estimates based on colors or the spectral energy distribution, consistent with the general behavior of spectroscopic line-strength diagnostics \citep{2007MNRAS.378..617K}. The practical limitation for highly reddened objects is instead observational: extinction may substantially reduce the signal-to-noise ratio, especially at shorter wavelengths, while interstellar extinction can render individual diagnostic features unsuitable for use.

\subsection{Metallicity dependence of the empirical calibrations}
\label{sec:metallicity_test}

The present calibrations are based on Galactic BA-type supergiants and therefore primarily represent the approximately solar-metallicity regime. Galactic BA supergiants generally show near-solar heavy-element abundances \citep{2006A&A...445.1099P,2012AA...543A..80F}. The available abundance analyzes for a subset of the calibration stars show no evidence for a distinct strongly $\alpha$-enhanced population \citep{2022AandA...668A..92W}. However, homogeneous determinations of $[\alpha/{\rm Fe}]$ are not available for the complete reference sample, so modest star-to-star abundance variations cannot be excluded.

We tested the possible metallicity effect directly on the published EWs of SMC A--type supergiants from \citet{1999ApJ...518..405V}. Three stars fall within the temperature domain of the present calibrations. Individual metal-line relations systematically overestimate their temperatures, with direct residuals of $\sim$ +(1.4--2.8)~kK. A differential comparison with the independent Galactic program star HD~222275 reduces the influence of the external temperature-scale zero point and still leaves positive residuals of $+0.8$--$1.6$~kK for the four single-line diagnostics that can be compared consistently, with a median offset of approximately $+1.2$~kK. The direction of this shift is consistent with the expected weakening of metal lines at lower abundances: when interpreted with a Galactic calibration, the reduced line strength is translated into an artificially higher $T_{\rm eff}$.

In contrast, selected line-ratio diagnostics show residuals comparable to the intrinsic scatter of the corresponding Galactic calibrations. In particular, ratios involving lines of the same ion appear substantially less sensitive to change in metallicity, because part of the abundance dependence is common to both lines. This behavior is qualitatively consistent with the expected advantage of differential line-strength indicators, although line ratios involving different elements may retain a dependence on the abundance pattern \citep{2019MNRAS.485.1310J}.

We therefore conclude that the individual metal-line relations presented here are composition-sensitive and should not be transferred directly to substantially metal-poor populations such as the SMC. The selected line ratios appear more robust, but their broader applicability requires validation with larger low-metallicity samples. The present SMC comparison should thus be regarded as an empirical test of the applicability limits of the Galactic calibration rather than as a universal metallicity correction.

\subsection{Program-star temperatures and comparison with literature}
\label{sec}

The retained empirical calibrations were applied to the program stars using only those indicators that passed the quality-control procedure and remained within the calibrated range of the corresponding relationship. For each star, the final $T_{\rm eff}$ was obtained by combining the individual temperature estimates retained after the consistency filtering described in Sect.~\ref{sec}.

The resulting empirical temperature scale comprises 84 program stars and is presented in Table~\ref{tab:program_teff}. For each object, the table lists the derived $T_{\rm eff}$, its adopted uncertainty, spectral classification, and, where available, the corresponding literature value and reference. Literature temperatures were adopted for 56 program stars, whereas no comparison value was available for the remaining 28 objects.

For the subsample with estimates from the available literature, a comparison was made between the $T_{\rm eff}$ derived in this work and the adopted reference values. The temperature difference was defined as

\begin{equation}
\Delta T_{\rm eff}
=
T_{\rm eff}^{\rm this\,work}
-
T_{\rm eff}^{\rm lit}.
\label{eq:delta_teff_lit}
\end{equation}

\begin{figure}[h!] 
\resizebox{1.0\hsize}{!} 
{\includegraphics{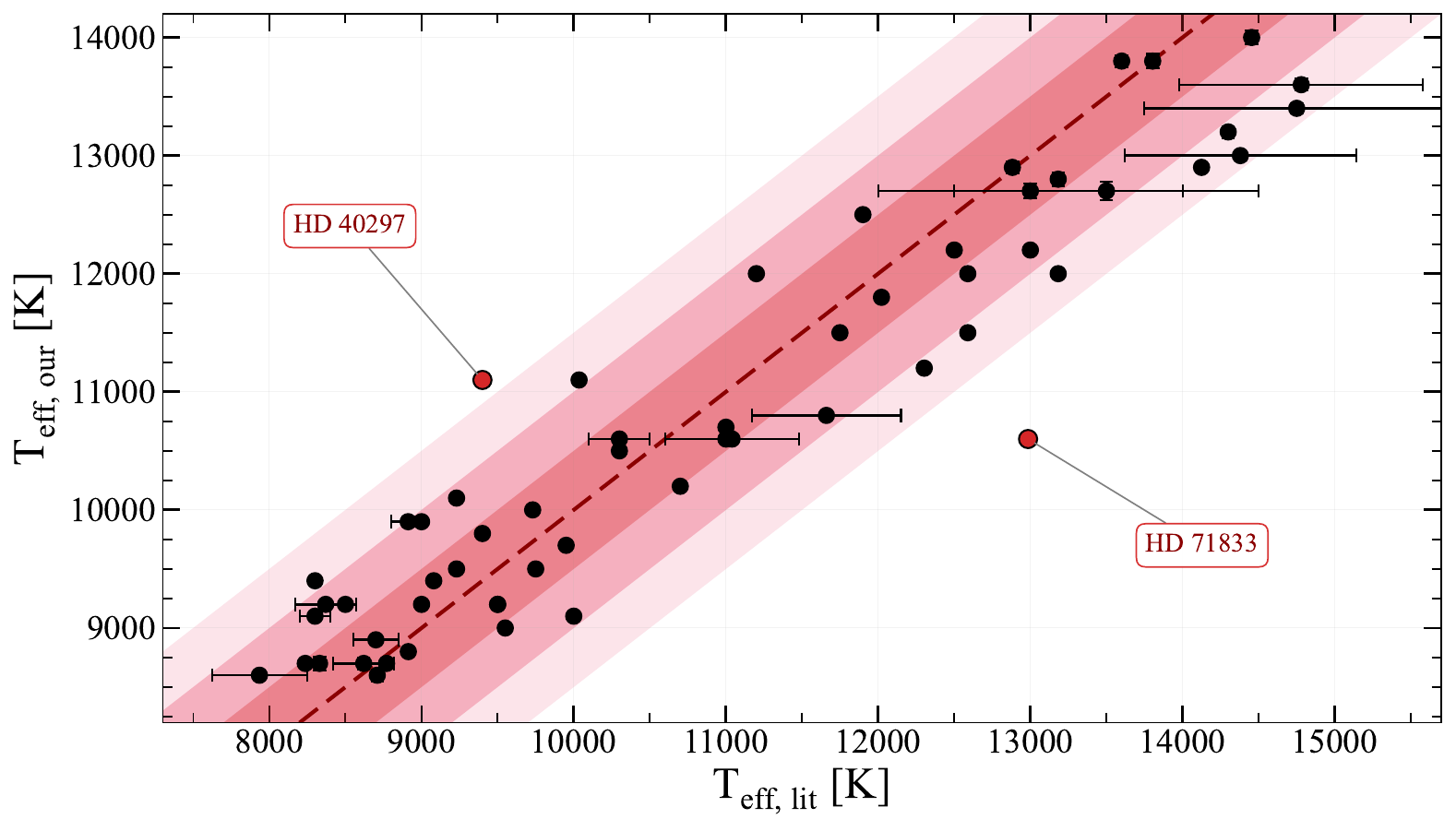}} 
\label{fig:teff_lit_comparison} \caption{
Comparison between the $T_{\rm eff}$ derived in this work and the adopted literature values. The dashed line indicates the one-to-one relationship. The shaded regions correspond to the $\pm1\sigma_{\Delta T}$ and $\pm2\sigma_{\Delta T}$ intervals. Red symbols mark the discrepant objects HD~40297 and HD~71833. Horizontal error bars indicate the reported uncertainties of the literature temperatures where available.
}
\end{figure}

The comparison is shown in Figure~\ref{fig:teff_lit_comparison}. The residual distribution has a mean value of $\langle\Delta T_{\rm eff}\rangle=-123$~K and a median value of $-168$~K. Thus, the empirical temperature scale does not show a substantial systematic offset relative to the values adopted from the literature.

Most objects follow the one-to-one relationship within the expected dispersion of an empirical comparison involving literature temperatures obtained with different observational material and analysis methods. In total, 54 out of 56 stars have $|\Delta T_{\rm eff}| \leq 1500$~K. Two objects show larger discrepancies and are highlighted in Figure~\ref{fig:teff_lit_comparison}.

The observed scatter should not be interpreted as the intrinsic uncertainty of the present empirical calibration alone because the values adopted in the literature do not represent a single homogeneous external temperature scale. Instead, this comparison provides an external consistency check for the empirical temperature scale across a heterogeneous literature sample. The most discrepant objects are discussed individually in Sect.~\ref{sec:individual_outliers}.

\subsection{Notes on stars with the largest temperature and other discrepancies}
\label{sec:individual_outliers}

\noindent\textit{HD~40297.}
For HD~40297, we derive $T_{\rm eff}=11\,100$~K, which is $1700$~K higher than the adopted literature value of $9400$~K from \citet{2014AJ....147..137L}. The difference may result from the use of different temperature diagnostics, since the present estimate is based on empirical EW- and line-depth relationships, whereas the literature value was obtained from an independent spectroscopic analysis. The distances used in the luminosity estimates were adopted from \citet{2025Univ...11..359V}. In our luminosity compilation, HD~40297 has $\log(L/L_\odot)=4.08\pm0.05$, placing it toward the lower-luminosity part of the BA-supergiant regime and close to the transition toward bright giants. In this regime, the selected line strengths may also be affected by atmospheric parameters other than $T_{\rm eff}$, particularly surface gravity and luminosity class. A dedicated atmospheric analysis is required to clarify the origin of the discrepancy.

\noindent\textit{HD~71833.}
For HD~71833, the derived temperature of $10\,600$~K is lower than the adopted value of $12\,985$~K by $2385$~K. The literature $T_{\rm eff}$ was estimated from Str\"omgren photometry using the Moon--Dworetsky calibration \citep{2011AandA...525A..97M}. In our luminosity compilation, this object, also known as HR~3345, has a luminosity of $\log(L/L_\odot)=2.67\pm0.02$, which is substantially below the luminosities expected for the BA supergiants represented by the calibration sample. Moreover, HD~71833 is classified as an HgMn star, and its chemical peculiarity may affect the strengths and depths of the metal and helium lines used in the present empirical relationships. Therefore, this object should not be considered a representative test of the calibration accuracy for chemically normal BA-type supergiants.

\noindent\textit{HD~167356.}
This star has been classified as A0Ia in \citet{1955ApJS....2...41M} and as A3II in \citet{1988mcts.book.....H}. However, its spectral type is erroneously listed as ApSi in SIMBAD. Our data support a luminous late-B/early-A classification for HD~167356, with $T_{\rm eff}=10\,400$~K, corresponding approximately to a B9/A0 spectral type. We derive $\log(L/L_\odot)=4.47\pm0.08$. This luminosity is consistent with the lower-luminosity part of the BA-supergiant regime and supports the interpretation of HD~167356 as a luminous evolved star rather than a chemically peculiar Ap star.

\subsection{Virtual Observatory Data Access and Services}
\label{sec:vo_service}

As introduced in Sect.~\ref{sec:obs}, the observational dataset has been integrated into the KazVO infrastructure. The underlying data publication framework is driven by the GAVO Data Center Helper Suite (DaCHS; \citealp{2014A&C.....7...27D}), which manages the backend database and protocol endpoints.

Access to the 1075 TCO \'echelle spectra is provided via the Simple Spectral Access Protocol (SSAP; \citealp{2012ivoa.spec.0210T}). Within the service metadata, the published spectra are classified as complying with calibration level~2. This ensures that external client applications correctly interpret them as fully wavelength-calibrated and continuum-normalized one-dimensional (1D) arrays. To maximize machine-readability and interoperability, table parameters and keywords are annotated using Unified Content Descriptors (UCD; \citealp{2019ivoa.spec.1007G}) and Unified Astronomy Thesaurus (UAT) controlled vocabularies. This allows standard Virtual Observatory (VO) tools, such as TOPCAT \citep{2011ascl.soft01010T} and Aladin \citep{Bonnarel2000}, to discover, visualize, and ingest the spectra.

Researchers can query the TCO hot supergiants collection programmatically via the global IVOA Registry, or interactively explore the dataset through the dedicated KazVO web interface\footnote{\url{https://vo.fai.kz/obs_data.php?path=/tco_hot_supergiants/q/web}}.

\section{SUMMARY}
\label{sec:summary}

This work presents an empirical framework for the $T_{\rm eff}$ characterization of Galactic B5--A5-type supergiants from optical spectra. The calibration relationships were constructed from a reference sample with data adopted from literature and link $T_{\rm eff}$ to equivalent widths, central line depths, and ratios of selected spectral features. Together, the resulting relationships span the interval from approximately 8\,400 to 14\,700~K, while the domain of applicability of each individual relationship is defined by the corresponding diagnostic and is specified in the calibration tables.

When applied to the program stars, the relationships yield a homogeneous empirical $T_{\rm eff}$ scale that is in good agreement with the determinations adopted from the literature. The comparison demonstrates that directly measured line parameters can provide a consistent basis for the temperature characterization of Galactic BA supergiants, including objects observed with heterogeneous spectroscopic material. The use of multiple
independent diagnostics, followed by consistency filtering, reduces the dependence of the final estimate on any individual spectral feature and allows the available information in an optical spectrum to be used more effectively.

The present relationships are not intended to replace detailed quantitative analyzes of stellar atmospheres. Their purpose is to provide an empirical temperature scale based on a restricted set of observable line parameters, whereas a full atmospheric analysis is required to determine the coupled effects of surface gravity, chemical composition, microturbulence, rotational broadening, stellar winds, and departures from LTE. The relationships should therefore be applied only within their specified validity ranges and with particular caution for stars with strong emission, shell components, chemically peculiar spectra, severe blending, or complex line profiles. Such objects may require an individual spectroscopic analysis beyond the scope of the present calibration framework.

Complete machine-readable calibration tables, including fitted coefficients, validity ranges, and uncertainty information, are publicly available together with a Python tool that implements the relationships and their uncertainty propagation \citep{Vaidman2026Zenodo}. In addition, the reduced continuum-normalized spectra and their associated metadata are published through the Kazakhstan National Virtual Observatory as a
VO-compliant spectral service. The combination of publicly available calibrations, software, and observational data provides a reproducible basis for further studies of Galactic hot supergiants and for the application of the presented framework to independent spectroscopic samples. In a follow up study, we will present luminosity-sensitive spectroscopic criteria for a similar group of hot BA supergiants.

\section{Data Availability}

The machine-readable calibration tables and a Python calculator implementing the empirical effective-temperature relationships are publicly available on Zenodo at \href{https://doi.org/10.5281/zenodo.20811472}{10.5281/zenodo.20811472}.

The spectral-line measurements were obtained with the \textsc{MERLIS} pipeline. The source code and associated configuration files are available at \url{https://github.com/nva1dman/merlis}.

The reduced continuum-normalized spectra and their associated metadata are available through the Kazakhstan National Virtual Observatory (KazVO) via the \href{https://vo.fai.kz/obs_data.php?path=/tco_hot_supergiants/q/web} {KazVO spectral service}.

The reduced OAN SPM spectra used in this study are available from the authors upon request.

\begin{acknowledgments}
This research was funded by the Science Committee of the Ministry of Science and Higher Education of the Republic of Kazakhstan (Grant No. AP23484898).
This research has made use of the resources, data, and/or
services of the Kazakhstan National Virtual Observatory (KazVO), supported by the
Fesenkov Astrophysical Institute and being a member of the IVOA.
This research has made use of the SIMBAD database, operated at CDS, Strasbourg, France; SAO/NASA ADS, ASAS, Gaia, data products.
This paper is partly based on observations obtained at the Canada-France-Hawaii Telescope (CFHT) which is operated by the National Research Council of Canada, the Institut National des Sciences de l$^{\prime}$Univers of the Centre National de la Recherche Scientifique de France.
SVZ acknowledges DGAPA-PAPIIT grant  IN105826.
\end{acknowledgments}

\begin{contribution}
N.L.V.: Software, Formal Analysis, Validation, Methodology, Data Curation, Visualization, Writing -- original draft, Writing -- review and editing. 
A.S.M.: Conceptualization, Methodology, Resources, Formal Analysis, Data Curation, Writing -- review and editing. 
S.T.N.: Formal Analysis, Validation, Writing -- review and editing. 
S.D.: Resources, Formal analysis.
S.A.K.: Data Curation, Writing -- review and editing. 
I.M.I.: Data Curation, Writing -- original draft.
C.T.O.: Writing -- review and editing. 
S.V.Z.: Resources, Data Curation.
D.E.Y.: Formal Analysis.
A.T.A.: Software.
A.A.K.: Project administration.
D.T.A.: Formal Analysis.
\end{contribution}

\software{astropy \citep{2013A&A...558A..33A,2018AJ....156..123A,2022ApJ...935..167A},  
SciPy \citep{2020NatMe..17..261V}, 
NumPy \citep{2020Natur.585..357H}, 
matplotlib \citep{2007CSE.....9...90H}
IRAF \citep{1986SPIE..627..733T,1993ASPC...52..173T}
          }

\begin{deluxetable*}{r l l r r r r @{\hspace{0.45cm}} r l l r r r r}
\tablecaption{Effective temperatures and spectral classifications of the program stars.
\label{tab:program_teff}}
\tablewidth{0pt}
\tablehead{
\colhead{No.} & \colhead{Star} & \colhead{Sp. type} & \colhead{$T_{\rm eff}$} & \colhead{$\sigma_T$} & \colhead{$T_{\rm eff,lit}$} & \colhead{Ref} &
\colhead{No.} & \colhead{Star} & \colhead{Sp. type} & \colhead{$T_{\rm eff}$} & \colhead{$\sigma_T$} & \colhead{$T_{\rm eff,lit}$} & \colhead{Ref} \\
\colhead{} & \colhead{} & \colhead{} & \colhead{(K)} & \colhead{(K)} & \colhead{(K)} & \colhead{} &
\colhead{} & \colhead{} & \colhead{} & \colhead{(K)} & \colhead{(K)} & \colhead{(K)} & \colhead{}
}
\startdata
1 & BD+60 51 & A2Iab & 9900 & 130 & 9000 & 1 & 43 & HD 40297 & A0Ib & 11100 & 120 & 9400 & 13 \\
2 & BD+61 153 & A0Ib & 9500 & 120 & 9750 & 1 & 44 & HD 40589 & A0Iab & 10800 & 120 & $11660\pm490$ & 7 \\
3 & BD+60 331 & B8Iab & 12000 & 120 & 13183 & 2 & 45 & HD 42400 & B5II & 13600 & 180 & -- & -- \\
4 & BD+60 333 & B5Ib & 14000 & 160 & 14454 & 2 & 46 & HD 43384 & B3Iab & 13600 & 150 & $14780\pm800$ & 7 \\
5 & BD+56 591 & A2Ia & 9400 & 120 & 9080 & 1 & 47 & HD 43820 & A2Ib & 10400 & 180 & -- & -- \\
6 & BD+43 1168 & B9Iab & 11000 & 120 & -- & -- & 48 & HD 43910 & A2Ia & 10200 & 130 & 10700 & 9 \\
7 & BD+62 2210 & B9Ia & 11500 & 140 & 11749 & 2 & 49 & HD 46769 & B7Ib-II & 12700 & 160 & $13000\pm1000$ & 14 \\
8 & BD+62 2313 & B7Ib & 13800 & 160 & 13804 & 2 & 50 & HD 46783 & B9Ib & 10700 & 120 & -- & -- \\
9 & BD+60 2542 & A1Iab & 9200 & 130 & 9000 & 1 & 51 & HD 47314 & B8Ib & 11000 & 120 & -- & -- \\
10 & HD 1070 & A4II & 9000 & 140 & -- & -- & 52 & HD 48452 & A7Ib & 9300 & 140 & -- & -- \\
11 & HD 2928 & A2Iab & 10000 & 120 & 9730 & 1 & 53 & HD 50064 & B5Ia & 12700 & 180 & $13500\pm1000$ & 15 \\
12 & HD 3940 & A1Ia & 10100 & 120 & 9230 & 1 & 54 & HD 55036 & A2Ib & 8900 & 180 & -- & -- \\
13 & HD 4717 & A2Iab & 9100 & 120 & 10000 & 1 & 55 & HD 55493 & A0/1Ia & 9800 & 160 & -- & -- \\
14 & HD 4841 & B5Ia & 13400 & 140 & $14750\pm1000$ & 3 & 56 & HD 58131 & B9Iab & 10500 & 130 & 10300 & 4 \\
15 & HD 5776 & A2Iab & 9200 & 120 & 9500 & 1 & 57 & HD 58439 & A2Ib/II & 8600 & 140 & $7935\pm312$ & 16 \\
16 & HD 7720 & B5II & 13200 & 150 & 14300 & 4 & 58 & HD 58764 & B9Ia & 10400 & 120 & -- & -- \\
17 & HD 9233 & A4Iab & 8900 & 140 & -- & -- & 59 & HD 59612 & A5/7Iab/II & 8700 & 150 & $8620\pm200$ & 10 \\
18 & HD 9311 & B5Ib & 12900 & 130 & 14125 & 2 & 60 & HD 62888 & B9.5Iab & 9300 & 120 & -- & -- \\
19 & HD 9811 & A6Ia & 8700 & 160 & 8330 & 4 & 61 & HD 67456 & A3Ib/II & 9400 & 130 & 8300 & 17 \\
20 & HD 10756 & B8Iab & 12000 & 120 & 12589 & 2 & 62 & HD 71833 & B8II & 10600 & 130 & 12985 & 18 \\
21 & HD 11577 & A0II: & 11100 & 140 & 10035 & 5 & 63 & HD 147084 & A5II & 9200 & 140 & $8370\pm200$ & 10 \\
22 & HD 11831 & A2Ia & 8900 & 130 & -- & -- & 64 & HD 161695 & A0Ib & 9700 & 120 & 9950 & 19 \\
23 & HD 13267 & B5Ia & 13000 & 140 & $14380\pm760$ & 7 & 65 & HD 167356 & A0Ia & 10400 & 120 & -- & -- \\
24 & HD 13717 & A0II: & 11300 & 120 & -- & -- & 66 & HD 167838 & B3Ia/ab & 13800 & 150 & 13600 & 20 \\
25 & HD 14010 & B9Ia & 11600 & 120 & -- & -- & 67 & HD 175687 & B9/A0Ib & 9800 & 120 & 9400 & 17 \\
26 & HD 14322 & B8Iab & 12200 & 120 & 13000 & 3 & 68 & HD 197345 & A2Ia & 8900 & 135 & $8700\pm150$ & 22 \\
27 & HD 14542 & B8Iab & 12200 & 120 & 12500 & 3 & 69 & HD 209900 & A0Ib & 9200 & 130 & 9500 & 1 \\
28 & HD 14899 & A0Ib & 10600 & 120 & 11000 & 3 & 70 & HD 211971 & A2Ib & 9500 & 120 & 9230 & 1 \\
29 & HD 15316 & A2Iab & 8700 & 130 & 8770 & 1 & 71 & HD 216927 & B9Ia & 11500 & 120 & 12589 & 2 \\
30 & HD 15497 & B6Ia & 12800 & 160 & 13183 & 2 & 72 & HD 222275 & A3II & 9200 & 130 & 8500 & 17 \\
31 & HD 15620 & B8Iab & 12000 & 130 & 11200 & 4 & 73 & HD 223767 & A5Ib & 8800 & 130 & 8913 & 2 \\
32 & HD 16778 & A1Ia & 9000 & 120 & 9550 & 2 & 74 & HD~228347$^\ast$ & A0Ib & 9400 & 140 & -- & -- \\
33 & HD 17086 & A5II & 8700 & 135 & 8236 & 8 & 75 & HD 236995 & A0Ib & 11800 & 120 & 12022 & 2 \\
34 & HD 17088 & B9Ia & 11200 & 120 & 12303 & 2 & 76 & HD 237153 & B8Ib & 13000 & 140 & -- & -- \\
35 & HD 17145 & B6Iab & 12900 & 150 & 12882 & 2 & 77 & HD 239886 & B8Ia & 10700 & 120 & -- & -- \\
36 & HD 17378 & A5Ia & 8600 & 150 & 8709 & 2 & 78 & HD 239895 & B8Ia & 12300 & 130 & -- & -- \\
37 & HD 17857 & B8Ib & 12500 & 130 & 11900 & 9 & 79 & HD 239950 & B8 & 9800 & 150 & -- & -- \\
38 & HD 21389 & A0Ia & 10600 & 120 & $11040\pm440$ & 7 & 80 & HD 248587 & A0Iab & 9500 & 130 & -- & -- \\
39 & HD 27381 & A4Ib & 8900 & 140 & -- & -- & 81 & HD 253250 & A0Iab & 9500 & 120 & -- & -- \\
40 & HD 28747 & A0II & 10600 & 120 & -- & -- & 82 & HD 332757 & A2Iab & 9200 & 120 & -- & -- \\
41 & HD 34578 & A5II & 9100 & 140 & $8300\pm100$ & 10 & 83 & HD~332996$^\ast$ & A0Ib & 9200 & 130 & -- & -- \\
42 & HD 35600 & B9Ib & 10700 & 120 & 11000 & 11 & 84 & HD~333030$^\ast$ & A0Ia & 10200 & 140 & -- & -- \\
\enddata
\tablecomments{
An asterisk (\(\ast\)) marks programme stars for which the OAN SPM spectra constitute the only observational material used in this work. The remaining stars are represented by spectra obtained at one or more of the TCO, ATO, and CFHT facilities. Spectral classifications were retrieved from SIMBAD. The listed effective temperatures correspond to the robust estimates obtained from the retained empirical calibrations. Literature temperatures and their uncertainties are left blank where no adopted literature value is provided. References: (1)~\citep{1999AandA...346..819V}; (2)~\citep{1992AandAS...94..211G}; (3)~\citep{1999AA...349..553M}; (4)~\citep{2010AN....331..349H}; (5)~\citep{2022AJ....163..152S}; (6)~\citep{2020MNRAS.492.2709P}; (7)~\citep{2009AandA...501..297Z}; (8)~\citep{2016AandA...591A.118S}; (9)~\citep{2008AandA...478..823M}; (10)~\citep{2010MNRAS.402.1369L}; (11)~\citep{2002PAOB..73..153G}; (12)~\citep{2012MNRAS.423.3268K}; (13)~\citep{2014AJ....147..137L}; (14)~\citep{2013AandA...557A.114A}; (15)~\citep{2010AandA...513L..11A}; (16)~\citep{1997PASP..109..958B}; (17)~\citep{1995ApJS...99..659V}; (18)~\citep{2011AandA...525A..97M}; (19)~\citep{2007MNRAS.374..664C}; (20)~\citep{1984AA...132..151L}; (21)~\citep{2022AJ....164..228L}; (22)~\citep{2012AA...543A..80F}.}
\end{deluxetable*}

\bibliography{sample701}{}
\bibliographystyle{aasjournalv7}
\end{document}